\documentclass[twoside]{article}
\usepackage{amsfonts}

\catcode`\@=11 \long\def\@makefntext#1{ \protect\noindent \hbox to
3.2pt {\hskip-.9pt
$^{{\eightrm\@thefnmark}}$\hfil}#1\hfill}       

\def\@makefnmark{\hbox to 0pt{$^{\@thefnmark}$\hss}}    

\def\ps@myheadings{\let\@mkboth\@gobbletwo
\def\@oddhead{\hfill\hbox{}\rightmark}
\def\@oddfoot{}\def\@evenhead{\leftmark\hbox{}\hfill}\def\@evenfoot{}
\def\sectionmark##1{}\def\subsectionmark##1{}}
\oddsidemargin=\evensidemargin 
\newcommand{\e}{\frac{e^{i(t'-t)(E_1-E_2)/\la^2}}{\la^2}}
\newcommand{\ga}{\gamma}
\newcommand{\la}{\lambda}
\newcommand{\om}{\omega}
\newcommand{\ep}{\varepsilon}
\newcommand{\dl}{\delta}
\newcommand{\ba}{\begin{array}{l}}
\newcommand{\ea}{\end{array}}
\font\tenrm=cmr10 \font\tenbf=cmbx10 \font\ninerm=cmr9
\font\eightrm=cmr8 \font\eightit=cmti8
\def\abstracts#1#2#3{{
    \centering{\begin{minipage}{4.5in}\footnotesize\baselineskip=10pt
    \parindent=0pt #1\par
    \parindent=15pt #2\par
    \parindent=15pt #3
    \end{minipage}}\par}}
\newcounter{sectionc}\newcounter{subsectionc}\newcounter{subsubsectionc}
\renewcommand{\section}[1] {\vspace{12pt}\addtocounter{sectionc}{1}
\setcounter{subsectionc}{0}\setcounter{subsubsectionc}{0}\noindent
    {\tenbf\thesectionc. #1}\par\vspace{5pt}}
\renewcommand{\subsection}[1] {\vspace{12pt}\addtocounter{subsectionc}{1}
    \setcounter{subsubsectionc}{0}\noindent
  {\bf\thesectionc.\thesubsectionc. {\kern1pt \bfit #1}}\par\vspace{5pt}}
\renewcommand{\subsubsection}[1] {\vspace{12pt}\addtocounter{subsubsectionc}{1}
    \noindent{\tenrm\thesectionc.\thesubsectionc.\thesubsubsectionc.
    {\kern1pt \tenit #1}}\par\vspace{5pt}}
\newcommand{\nonumsection}[1] {\vspace{12pt}\noindent{\tenbf #1}
    \par\vspace{5pt}}
\newcounter{appendixc}
\renewcommand{\appendix}[1] {\vspace{12pt}
        \refstepcounter{appendixc}
        \setcounter{lemma}{0}
        \setcounter{theorem}{0}
        \setcounter{equation}{0}
        \renewcommand{\theappendixc}{\Alph{appendixc}}
        \renewcommand{\thelemma}{\Alph{appendixc}.\arabic{lemma}}
        \renewcommand{\thetheorem}{\Alph{appendixc}.\arabic{theorem}}
        \renewcommand{\theequation}{\Alph{appendixc}.\arabic{equation}}
        \noindent{\tenbf Appendix#1}\par\vspace{5pt}}

\renewenvironment{thebibliography}[1]
    {\frenchspacing
     \ninerm
     \baselineskip=11pt
     \begin{list}{\arabic{enumi}.}
        {\usecounter{enumi}\setlength{\parsep}{0pt}
     \setlength{\leftmargin 12.7pt}{\rightmargin 0pt} 
         \setlength{\itemsep}{0pt} \settowidth
    {\labelwidth}{#1.}\sloppy}}{\end{list}}
\def\eightcirc{
\begin{picture}(0,0)
\put(4.4,1.8){\circle{6.5}}
\end{picture}}
\def\eightcopyright{\eightcirc\kern2.7pt\hbox{\eightrm c}}

\newcommand{\copyrightheading}[1]
    {\vspace*{-2.5cm}\baselineskip=10pt{\flushleft
    {\footnotesize Infinite Dimensional Analysis, Quantum Probability
          and Related Topics #1}\\
    {\footnotesize\copyright\kern2pt World Scientific Publishing
     Company}\\
     }}
\def\keywords#1{{
    \centering{\begin{minipage}{4.5in}\footnotesize\baselineskip=10pt
    {\footnotesize\it Keywords}\/: #1
     \end{minipage}}\par}}
\newtheorem{theorem}{Theorem}[sectionc] 
\newtheorem{lemma}{Lemma}[sectionc]

\def\qed{\hbox{${\vcenter{\vbox{         
   \hrule height 0.4pt\hbox{\vrule width 0.4pt height 6pt
   \kern5pt\vrule width 0.4pt}\hrule height 0.4pt}}}$}}

\def\theequation{\thesectionc.\arabic{equation}}  
\begin{document}
\setlength{\textheight}{7.7truein}    

\markboth{\protect{\footnotesize\it A Stochastic Golden Rule for
the LDL}}{\protect{\footnotesize\it A Stochastic Golden Rule for
the LDL}}

\normalsize\baselineskip=13pt\thispagestyle{empty}
\setcounter{page}{1}

\copyrightheading{}

\vspace*{1in}

 \centerline{\tenbf A STOCHASTIC GOLDEN RULE} \baselineskip=13pt
 \centerline{\tenbf AND QUANTUM LANGEVIN EQUATION}\baselineskip=13pt
 \centerline{\tenbf FOR THE LOW DENSITY LIMIT}

\vspace*{0.37truein}

\centerline{\eightrm L. ACCARDI\footnote{E-mail:
accardi@volterra.mat.uniroma2.it}}
\baselineskip=12pt\centerline{\eightit Centro Vito Volterra,
Universita di Roma Tor Vergata 00133,Italia}

\vspace*{10pt}

\centerline{\eightrm A. N. PECHEN\footnote{E-mail:
pechen@mi.ras.ru}\,\,\,  and I. V. VOLOVICH\footnote{E-mail:
volovich@mi.ras.ru}}

 \baselineskip=12pt\centerline{\eightit Steklov Mathematical Institute, Russian Academy of Sciences,}
 \baselineskip=10pt\centerline{\eightit Gubkin St. 8, GSP-1, 117966, Moscow, Russia}

\vspace*{1truein}

\abstracts{\eightrm A rigorous derivation of quantum Langevin equation from microscopic dynamics
in the low density limit is given. We consider a quantum model of a microscopic system (test
particle) coupled with a reservoir (gas of light Bose particles) via interaction of scattering
type. We formulate a mathematical procedure (the so-called stochastic golden rule) which allows us
to determine the quantum Langevin equation in the limit of large time and small density of
particles of the reservoir. The quantum Langevin equation describes not only dynamics of the system
but also the reservoir. We show that the generator of the corresponding master equation has the
Lindblad form of most general generators of completely positive semigroups.}{}{}

\vspace*{10pt} \keywords{low density, quantum stochastic
equations, quantum Langevin equation}

\vspace*{4pt} \baselineskip=13pt \normalsize\tenrm
\section{Introduction}
\noindent \looseness1 In the past years many important physical models have been investigated by
using the stochastic limit method (see \cite{ALV,AKV} for a survey). In particular Fermi golden
rule has been generalized to a {\it stochastic golden rule} which allows to solve the following
problem: give a quantum Hamiltonian system, determine the associated Langevin equation (including
the structure of the so-called master fields or quantum noises) and obtain as a corollary the
master equation associated to this system.

The Langevin equation is an asymptotic equation, valid in a regime of large times and
small parameters (e.f. weak coupling) and its interest for the applications comes from
the fact that the relevant physical phenomena in this regime are much more transparent
than in the original Hamiltonian equation and in this sense an approximate equation gives
a better description of the physical phenomenon than the original exact equation.

The stochastic golden rule plays a fundamental role in the stochastic limit because it
allows to separate the problem of finding the correct solution from the problem of giving
a rigorous proof of the convergence of the method.

Although well developed in the weak coupling case, in the low
density case, the stochastic golden rule was formulated
in~\cite{ALV} only for Fock fields and this limitation exclude of
the most interesting cases for the applications.

The development of a stochastic golden rule for the low density
limit required a totally different approach from that developed
for the weak coupling case and the first step in this direction
was done in the paper~\cite{apv} where the problem was solved
under the (very strong) assumption of rotating wave approximation
(RWA). In the present paper we bring to a completion the programme
begun in~\cite{apv} and we establish the stochastic golden rule
for general discrete spectrum system interacting with a Boson
reservoir without RWA. This also gives, when combined with the
analytical estimates of part III of~\cite{ALV}, a simplified proof
of the deduction of the stochastic Schr\"odinger equation for the
low density limit. The new proof is much simpler and allows a
deeper understanding of the emergence of the scattering operator,
more precisely the $T$-operator, in this equation.

To describe a quantum physical model to which we apply the low
density limit let us consider an N-level atom immersed in a free
gas whose molecules can collide with the atom; the gas is supposed
to be very dilute. Then the reduced time evolution for the atom
will be Markovian, since the characteristic time  $t_S$  for
appreciable action of the surroundings on the atom (time between
collisions) is much larger than the characteristic time  $t_R$ for
relaxation of correlations in the surroundings. Rigorous results
substantiating this idea have been obtained in~\cite{dumcke}.

The dynamics of the N-level atom interacting with the free gas
converges, in the low density limit, to the solution of a quantum
stochastic differential equation driven by quantum Poisson
noise~\cite{AcLu}. Indeed, from a semiclassical point of view,
collision times, being times of occurrence of rare events, will
tend to become Poisson distributed, whereas the effect of each
collision will be described by the (quantum-mechanical) scattering
operator of the atom with one gas particle (see the description of
the quantum Poisson process in~\cite{kumerrer}).

We consider a microscopic system (test particle) interacting with
a gas of Bose particles (reservoir) via interaction of scattering
type so that the particles of the reservoir are only scattered and
not created or destroyed. The reservoir is supposed to be a very
dilute gas at equilibrium state. We obtain starting from
microphysical dynamics, in the limit of long time and small
density of Bose particles, the quantum Langevin equation which
describes the evolution of any system observable. The Langevin
equation includes not only the system but also the reservoir
dynamics. After averaging over the reservoir equilibrium state we
obtain a master equation which describes the evolution of any
system observable, i.e. the reduced evolution of the test
particle. The generator of this master equation has the standard
Lindblad form~\cite{lindblad}.

Starting from the quantum stochastic equation~(\ref{normordeq}) for the evolution operator one can
easy deduce the corresponding quantum linear Boltzmann equation for the density matrix of the
system. This is a linear dissipative equation describing an irreversible evolution of the test
particle. A general structure of a master equation driving a completely positive time evolution for
a test particle in a quantum (Rayleigh) gas has been investigated in~\cite{VL}. It turns out that
general requirements of the underlying symmetry are not sufficient to fix the form of the master
equation and one needs a microscopic derivation of the master equation~\cite{V}. A linear
dissipative equation describing the long time dynamics of a particle interacting with a reservoir
(a quantum Fokker-Planck equation) has been studied in~\cite{qfp} where the problem of rigorous
derivation of such equations was emphasized.

Using the white noise approach developed in~\cite{ALV} and the
energy representation~\cite{apv} we derive a new algebra for the
quadratic master fields in the low density limit. An advantage of
this method is the simplicity of derivation of the white noise
equations and in the computation of correlation functions. We show
that the drift term in the QSDE is simply related with
$T$-operator describing the scattering of the system on the one
particle sector of the reservoir.

The main results of the paper are:

(i) the explicit determination of the algebra of the master field in energy representation
(section 3)

(ii) the determination of the white noise Hamiltonian equation~(\ref{equ}) for the low density
limit

(iii) the determination of the (causally) ordered
form~(\ref{normordeq}) of this equation which, as known from the
general theory developed in~\cite{ALV}, is equivalent to a quantum
stochastic differential equation in the sense of~\cite{HP}

(iv) formula~(\ref{drift}), which gives an explicit expression for
the drift in terms of the 1-particle $T$-operator

(v) the determination of the quantum Langevin
equation~(\ref{langevin}) and the quantum Markovian
generator~(\ref{QMG}) of the corresponding master equation.

Our strategy is the following. In section~2 we describe the model
which we consider; and for this model, we introduce the
Fock-anti-Fock representation. In section~3 we introduce the
rescalled fields whose stochastic limits are the master fields in
the LDL. In section~4 we derive the commutation relations of the
master fields (Theorem~1). Then in section~5 we construct a
concrete representation of this algebra. Using this representation
we write in section~6 the stochastic Schr\"odinger
equation~(\ref{equ}) for the limiting evolution operator. In
section~7 we rewrite the stochastic Schr\"odinger equation in
normally ordered form, i.e. in the form of a quantum stochastic
differential equation. In section~8 we prove formula~(\ref{drift})
which expresses the drift, in the equation for the evolution
operator in the LDL in terms of: the 1-particle $T$-operator, the
1-particle gas Hamiltonian and the spectral projections of the
system Hamiltonian. In section~9 we obtain the quantum Langevin
equation and the quantum Markovian generator of the corresponding
master equation.


\section{The Model}

Now let us explain our notations.
Let $ {\cal H}_{\rm S} $ be a Hilbert space of the system with the Hamiltonian
$ H_{\rm S} $. We will suppose that the system Hamiltonian $H_{\rm S}$ has a discrete
spectrum
$$
 H_{\rm S}=\sum\limits_k\ep_kP_k
$$
where $\ep_k$ are the eigenvalues and $P_k$ the spectral projections.

We will consider the case of a Boson reservoir in this paper.
Therefore the reservoir is described by the Boson Fock space
$\Gamma({\cal H}_1)$ over the one particle Hilbert space ${\cal
H}_1=L^2(\mathbb R^d)$ (the scalar product in ${\cal H}_1$ we
denote as $<\cdot,\cdot>$), where $d=3$ in physical case.
Moreover, the free Hamiltonian of the reservoir is given by
$H_R:=\Gamma(H_1)$ (the second quantization of the one particle
Hamiltonian $ H_1 $) and the total Hamiltonian of the compound
system is given by a self--adjoint operator on the total Hilbert
space ${\cal H}_S\otimes \Gamma({\cal H}_1)$, which has the form
$$
 H_{\rm tot}:=H_S\otimes1+1\otimes H_R+V=:H_0+V
$$
Here $V$  is the interaction Hamiltonian between the
system and the reservoir.
The evolution operator at time $t$ is given by:
\begin{equation}\label{EvOp}
 U(t):=e^{itH_0}\cdot e^{-itH_{\rm tot}}
\end{equation}
and it satisfies the differential equation
$$
 \partial_tU(t)=-iV(t)U(t)
$$
where the quantity the evolved interaction $V(t)$ is defined by
$$
 V(t)=e^{itH_0}Ve^{-itH_0}.
$$

The interaction Hamiltonian will be assumed to have the following
form:
$$
 V= D\otimes A^+(g_0)A(g_1)+D^+\otimes A^+(g_1)A(g_0)
$$
where $ D $ is a bounded operator in $ {\cal H}_S $,
$ D\in {\bf B}({\cal H}_S) $, $ A $ and $ A^+ $ are annihilation
and creation operators and $ g_0, g_1\in {\cal H}_1$
are form-factors describing the interaction of the
system with the reservoir. Therefore, with the notations
$$
 S_t:=e^{itH_1} \ ;\qquad D(t):=e^{itH_S}De^{-itH_S}
$$
the evolved interaction can be written in the form
\begin{equation}\label{Hint}
 V(t)= D(t)\otimes A^+(S_tg_0)A(S_tg_1)+D^+(t)\otimes
 A^+(S_tg_1)A(S_tg_0)
\end{equation}

The initial state of the compound system is supposed to be of the
form
$$
 \rho = \rho_{\rm S}\otimes\varphi^{(\xi)}.
$$
Here $\rho_{\rm S}$ is arbitrary density matrix of the system and
the initial state of the reservoir $\varphi^{(\xi)}$ is the Gibbs
state, at inverse temperature $\beta$, of the free evolution, i.e.
the gauge invariant quasi-free state characterized by
\begin{equation}\label{state}
 \varphi^{(\xi)}(W(f))=\exp\Bigl(-{1\over 2}<f,(1+\xi e^{-\beta H_1})
 (1-\xi e^{-\beta H_1})^{-1}f>\Bigr)
\end{equation}
for each $f\in {\cal H}_1$. Here $ W(f) $ is the Weyl operator,
$\beta$ the inverse temperature of the reservoir,
$\xi=e^{\beta\mu}$ the fugacity, $\mu$ the chemical potential. We
suppose that the temperature $\beta^{-1}>0$. Therefore for
sufficiently low density one is above the transition temperature,
and no condensate is present.

We will study the dynamics, generated by the
Hamiltonian~(\ref{Hint}) and the initial state of the
reservoir~(\ref{state}) in the low density regime: $n\to 0$,
$t\sim 1/n$ ($n$ is the density of particles of the reservoir). We
do not fix the initial state of the system so our results can be
applied to an arbitrary initial density matrix of the system. In
the low density limit the fugacity $\xi$ and the density of
particles of the reservoir $n$ have the same asymptotic, i.e.
$$
 \lim\limits_{n\to 0}\frac{\xi(n)}{n}=1
$$
Therefore the limit $n\to 0$ is equivalent to the limit $\xi\to 0$.

Throughout this paper, for simplicity, the following  technical
condition is assumed: the two test functions in the interaction
Hamiltonian have disjoint supports in the energy representation.
This condition is invariant under the action of any function of
$H_1$ and implies that the two test function $g_0, g_1$ in the
interaction Hamiltonian satisfy:
$$
 <g_0,S_te^{-\beta H_1}g_1>=0\qquad \forall t\in{\mathbb R}.
$$

In the paper~\cite{apv} we obtained for the model described above a quantum stochastic
differential equation under additional rotating wave approximation condition:
\begin{equation}\label{rwa}
 D(t)=e^{itH_S}De^{-itH_S}=e^{-it\omega_0}D
\end{equation}
for some real number $\omega_0$.

In the present article we will derive the white noise Schr\" odinger equation without
assuming any relation between $D$ and $H_{\rm S}$.

Let us rewrite the free evolution of $D$ in a form which is convenient for derivation of
the white noise equation. For this we introduce the set of all Bohr frequencies
$$
 {\rm B(H_S)}=:{\rm B}:=
  \{\omega\, |\, \exists\, \ep_k,\ep_m\in {\rm spec\, H_S}\ {\rm s.t.}\ \omega=\ep_k-\ep_m\}
$$
Using this notion and the properties of the spectral projections one can rewrite the free
evolution of an arbitrary system operator $D$ in the form
$$
 D(t)=\sum\limits_{k,m}e^{it(\ep_k-\ep_m)}P_kDP_m=\sum\limits_{\omega\in {\rm B}}
  e^{-it\omega}D_\omega
$$
where we denote
$$
 D_\omega:=\sum\limits_{k,m: \ep_m-\ep_k=\omega}P_kDP_m
$$

We realize the representation space
as the tensor product of a Fock and anti-Fock representations.
Then the expectation values with respect to the state
$ \varphi^{(\xi)} $ for the
model with the interaction Hamiltonian~(\ref{Hint}) can be
conveniently represented as the vacuum expectation values in the
Fock-anti-Fock representation for the modified Hamiltonian.

Denote by ${\cal H}_1^\iota$ the conjugate of ${\cal H}_1$, i.e. ${\cal H}_1^\iota$ is
identified to ${\cal H}_1$ as a set and the identity operator
$\iota :\ {\cal H}_1\longrightarrow {\cal H}_1$ is antilinear
$$
 \iota(\la f):=\bar \la \iota (f)
$$
$$
 <\iota(f),\iota(g)>_\iota \ :=\ <g,f>
$$
Then, ${\cal H}_1^\iota $ is a Hilbert space and, if the vectors of ${\cal H}_1$ are thought
as ket-vectors $|\xi>$, then the vectors of ${\cal H}_1^\iota$ can be thought as bra-vectors
$<\xi|$. The corresponding Fock space $ \Gamma({\cal H}_1^\iota) $ is called
the anti-Fock space.

It was shown in~\cite{AcLu} that, with the notations $
D_{\omega,0}=D_\omega,\, D_{\omega,1}=D_\omega^+ $ and
$\la=+\sqrt{\xi}$, the part of the modified Hamiltonian which
gives a nontrivial contribution in the LDL, acts in $ {\cal
H}_S\otimes\Gamma({\cal H}_1)\otimes \Gamma({\cal H}_1^\iota) $ as
$$
 H_\la(t)=\sum\limits_{\ep=0,1}\sum\limits_{\omega\in B}
 D_{\omega,\ep}e^{-(-1)^\ep it\omega}\otimes\Bigl(A^+(S_tg_\ep)A(S_tg_{1-\ep})\otimes 1
$$
$$
 +\la\bigl(A(S_tg_{1-\ep})\otimes
 A(S_tLg_\ep)+A^+(S_tg_\ep)\otimes A^+(S_tLg_{1-\ep})\bigr)\Bigr).
$$
Here $ A $ and $ A^+ $ are Bose annihilation and creation operators
acting in the Fock spaces $ \Gamma({\cal H}_1) $ and
$ \Gamma({\cal H}_1^\iota) $ and $ L:=e^{-\beta H_1/2} $.

The interaction Hamiltonian $H_\la(t)$ determines the evolution operator $ U_t^{(\la)} $
which is the solution of the Schr\"odinger equation
in interaction representation:
$$
\partial_tU_t^{(\la)}=-iH_\la(t)U_t^{(\la)}
$$
with initial condition
$$
U_0^{(\la)}=1.
$$
This is equivalent to the following integral equation for the
evolution operator
$$
U_t^{(\la)}=1-i\int\limits_0^tdt'H_\la(t')U_{t'}^{(\la)}.
$$

The convenience of the Fock-anti-Fock representation, as it was mentioned above, is in the
fact that the expectation value, for example, of any time evolved system observable
$X_t=U^*(t)(X\otimes 1)U(t)$ with respect to the state~(\ref{state}) is equal to the vacuum
expectation value in the Fock-anti-Fock representation, i.e.
$$
 \varphi^{(\xi)}(U^*(t)(X\otimes 1)U(t))=
 <U^{*(\la)}_t(X\otimes 1\otimes 1)U^{(\la)}_t>_{\rm vac}
$$


\section{Energy Representation}
\noindent We will investigate the limit of the evolution operator
$U^{(\la)}_{t/\xi}$ when $ \xi\to +0 $ after the time rescaling $
t\to t/\xi $, where $ \xi=\la^2 $. After this time rescaling the
equation for the evolution operator becomes
$$
 \partial_tU_{t/\la^2}^{(\la)}=-i\sum\limits_{\omega\in B}\sum\limits_{\ep=0,1}D_{\omega,\ep}
 \otimes\bigl[N_{\ep,1-\ep,\la}((-1)^\ep\omega,t)
$$
$$
 +B_{1-\ep,\ep,\la}((-1)^{1-\ep}\omega,t)
 +B^+_{\ep,1-\ep,\la}((-1)^\ep\omega,t)\bigr]U^{(\la)}_{t/\la^2}
$$
where we introduced for each $\ep_1,\ep_2=0,1$ and $\om\in B$ the
rescaled fields:
\begin{equation}\label{dfn}
 N_{\ep_1,\ep_2,\la}(\omega,t):=\frac{1}{\la^2}e^{-it\omega/\la^2}A^+(S_{t/\la^2}g_{\ep_1})
 A(S_{t/\la^2}g_{\ep_2})\otimes 1
\end{equation}
\begin{equation}\label{dfb+}
 B_{\ep_1,\ep_2,\la}(\omega,t):=\frac{1}{\la}e^{it\omega/\la^2}
 A(S_{t/\la^2}g_{\ep_1})\otimes A(S_{t/\la^2}Lg_{\ep_2}).
\end{equation}
and $B^+_{\ep_1,\ep_2,\la}(\om,t)$ is the adjoint of
$B_{\ep_1,\ep_2,\la}(\om,t)$.

The energy representation for the creation and annihilation
operators is defined by the formulae
\begin{equation}\label{AEdef}
 A^+_E(g):=A^+(P_Eg),\qquad A_E(g):=A(P_Eg)
\end{equation}
with
$$
 P_E:=\frac{1}{2\pi}\int\limits_{-\infty}^{\infty}dtS_te^{-itE}=
 \dl(H_1-E)\eqno{(\ref{AEdef}a)}
$$
It has the properties
$$
 S_t=\int dEP_Ee^{itE}\eqno{(\ref{AEdef}b)}
$$
$$
 \vphantom{\int}P_EP_{E'}=\dl(E-E')P_E,\qquad P^*_E=P_E
$$

The meaning of the $\delta$-function in~(\ref{AEdef}a) is
explained in~\cite{ALV}, section~(1.2). In our case for $ {\cal
H}_1=L^2({\mathbb R}^d) $ the one-particle Hamiltonian is the
multiplication operator by the function $ \omega(k) $ and acts on
an element $ f\in L^2({\mathbb R}^d) $ as $
(H_1f)(k)=\omega(k)f(k)$ so that $P_E=\dl(\om(k)-E)$.

It is easy to check that
$$
 [A_E(f),A_{E'}^+(g)]=\dl(E-E')<f,P_Eg>.
$$

Using the energy representation~(\ref{AEdef}b), (\ref{dfn})
and (\ref{dfb+}) become respectively
$$
 N_{\ep_1,\ep_2,\la}(\omega,t)=\int\int dE_1dE_2N_{\ep_1,\ep_2,\la}(E_1,E_2,\omega,t)
$$
$$
 B_{\ep_1,\ep_2,\la}(\omega,t)=\int\int dE_1dE_2B_{\ep_1,\ep_2,\la}(E_1,E_2,\omega,t)
$$
where the rescaled fields in energy representation are given by
\begin{equation}\label{N}
 N_{\ep_1,\ep_2,\la}(E_1,E_2,\omega,t):=\frac{e^{it(E_1-E_2-\omega)/\la^2}}{\la^2}
 A^+_{E_1}(g_{\ep_1})A_{E_2}(g_{\ep_2})\otimes 1
\end{equation}
\begin{equation}\label{B}
 B_{\ep_1,\ep_2,\la}(E_1,E_2,\om,t):=
 \frac{e^{it(E_2-E_1+\omega)/\la^2}}{\la}A_{E_1}(g_{\ep_1})\otimes A_{E_2}(Lg_{\ep_2}).
\end{equation}
and $B^+_{\ep_1,\ep_2,\la}(E_1,E_2,\om,t)$ is the adjoint of
$B_{\ep_1,\ep_2,\la}(E_1,E_2,\om,t)$. Let us also denote
$$
 \ga_\ep(E):=\int\limits^0_{-\infty}<g_\ep,S_tg_\ep>e^{-itE}dt.
$$


\section{The Algebra of the Master Fields}
\noindent In this section we derive the algebra of master fields
which is the limit as $\la\to 0$ of rescaled
fields~(\ref{N}),~(\ref{B}) in the sense of convergence of
correlators.

Denote $\Omega$ the lattice of all linear combinations of the Bohr frequencies
with integer coefficients:
$$
 \Omega=\{\omega\,|\,\omega=\sum\limits_{k=1}^Nn_k\omega_k\ {\rm with}\ N\in\mathbb
 N,\, n_k\in\mathbb Z,\, \omega_k\in{\rm B}\}
$$
and extend the definitions~(\ref{N}) and~(\ref{B}), of the rescaled fields,
by allowing in them an arbitrary $\om\in\Omega$. The following theorem
describes the algebra of commutation
relations for the master field in the LDL.
\begin{theorem}\label{mastfi}
The limits of the rescaled fields
$$
 X_{\ep_1,\ep_2}(E_1,E_2,\omega,t):=
 \lim\limits_{\la\to 0}X_{\ep_1,\ep_2,\la}(E_1,E_2,\omega,t), \qquad X=B,B^+,N
$$
exist in the sense of convergence of correlators and satisfy the commutation relations
$$
 \vphantom{\e}
 [B_{\ep_1,\ep_2}(E_1,E_2,\omega,t),B^+_{\ep_3,\ep_4}(E_3,E_4,\omega',t')]=
 2\pi\dl_{\omega,\omega'}\dl_{\ep_1,\ep_3}\dl_{\ep_2,\ep_4}\dl(t'-t)
$$
\begin{equation}\label{crmf1}
 \vphantom{\e}
 \times\dl(E_1-E_3)\dl(E_2-E_4) \dl(E_1-E_2-\omega)
 <g_{\ep_1},P_{E_1}g_{\ep_1}><g_{\ep_2},P_{E_2}L^2g_{\ep_2}>
\end{equation}
$$
 \vphantom{\e}
 [B_{\ep_1,\ep_2}(E_1,E_2,\omega,t),N_{\ep_3,\ep_4}(E_3,E_4,\omega',t')]=
 2\pi\dl_{\ep_1,\ep_3}\dl(t'-t)
$$
\begin{equation}\label{crmf2}
 \vphantom{\e}
 \times\dl(E_1-E_3)\dl(E_1-E_2-\omega)
 <g_{\ep_1},P_{E_1}g_{\ep_1}>B_{\ep_4,\ep_2}(E_4,E_2,\omega-\omega',t')
\end{equation}
$$
 \vphantom{\e}
 [N_{\ep_1,\ep_2}(E_1,E_2,\omega,t),N_{\ep_3,\ep_4}(E_3,E_4,\omega',t')]=
 2\pi\dl(t'-t)
$$
$$
 \vphantom{\e}
 \times\Bigl\{\dl_{\ep_2,\ep_3}\dl(E_2-E_3)\dl(E_3-E_1+\omega)<g_{\ep_2},P_{E_2}g_{\ep_2}>
 N_{\ep_1,\ep_4}(E_1,E_4,\omega+\omega',t')
$$
\begin{equation}\label{crmf3}
 \vphantom{\e}
 -\dl_{\ep_1,\ep_4}\dl(E_1-E_4)\dl(E_3-E_1-\omega')<g_{\ep_1},P_{E_1}g_{\ep_1}>
 N_{\ep_3,\ep_2}(E_3,E_2,\omega+\omega',t)\Bigr\}
\end{equation}
The causal commutation relations of the master field are obtained
replacing in (\ref{crmf1}),(\ref{crmf2}),(\ref{crmf3}) the factor
$ \dl(t'-t) $ by $ \dl_+(t'-t) $ where the causal $\dl$-function
$\dl_+(t'-t)$ is defined in section (8.4) of~\cite{ALV}, $
2\pi\dl(E_1-E_2-\omega) $ by $ (i(E_1-E_2-\om-i0))^{-1} $ and
$2\pi\dl(E_3-E_1\pm\omega)$ by $ (i(E_3-E_1\pm\om-i0))^{-1} $.
\end{theorem}

\noindent {\bf Proof.} Introduce the operators:
$$
 N^i_{\ep_1,\ep_2,\la}(E_1,E_2,\omega,t):=
 \frac{e^{it(E_2-E_1-\omega)/\la^2}}{\la^2}
 \phantom{i}1\otimes A^+_{E_1}(Lg_{\ep_1})A_{E_2}(Lg_{\ep_2})
$$
with $A_E(Lg_\ep)$ and $A^+_E(Lg_\ep)$ defined by (\ref{AEdef}),
the commutators of the rescalled fields are:
$$
 [B_{\ep_1,\ep_2,\la}(E_1,E_2,\omega,t),B^+_{\ep_3,\ep_4,\la}(E_3,E_4,\omega',t')]=
 \frac{e^{i(t'-t)(E_1-E_2-\omega)/\la^2}}{\la^2}
$$
$$
 \vphantom{\e}
 \times\Bigl(\dl_{\ep_1,\ep_3}\dl_{\ep_2,\ep_4}e^{it'(\omega-\omega')/\la^2}\dl(E_1-E_3)\dl(E_2-E_4)
 <g_{\ep_1},P_{E_1}g_{\ep_1}><g_{\ep_2},P_{E_2}L^2g_{\ep_2}>
$$
$$
 \vphantom{\e}
 +\la^2\dl_{\ep_1,\ep_3}\dl(E_1-E_3)<g_{\ep_1},P_{E_1}g_{\ep_1}>
 N^i_{\ep_4,\ep_2,\la}(E_4,E_2,\omega'-\omega,t')
$$
\begin{equation}\label{cr1}
\vphantom{\e}
 +\la^2\dl_{\ep_2,\ep_4}\dl(E_2-E_4)<g_{\ep_2},P_{E_2}L^2g_{\ep_2}>
 N_{\ep_3,\ep_1,\la}(E_3,E_1,\omega'-\omega,t')\Bigr)
\end{equation}
$$
 [B_{\ep_1,\ep_2,\la}(E_1,E_2,\omega,t),N_{\ep_3,\ep_4,\la}(E_3,E_4,\omega',t')]=
 \frac{e^{i(t'-t)(E_1-E_2-\omega)/\la^2}}{\la^2}
$$
\begin{equation}\label{cr2}
 \vphantom{\e}
 \times\dl_{\ep_1,\ep_3}\dl(E_1-E_3)<g_{\ep_1},P_{E_1}g_{\ep_1}>
 B_{\ep_4,\ep_2,\la}(E_4,E_2,\omega-\omega',t')
\end{equation}
$$
 \vphantom{\e}
 [N_{\ep_1,\ep_2,\la}(E_1,E_2,\omega,t),N_{\ep_3,\ep_4,\la}(E_3,E_4,\omega',t')]
$$
$$
 =\frac{e^{i(t'-t)(E_3-E_1+\omega)/\la^2}}{\la^2}
 \dl_{\ep_2,\ep_3}\dl(E_2-E_3)<g_{\ep_2},P_{E_2}g_{\ep_2}>
 N_{\ep_1,\ep_4,\la}(E_1,E_4,\omega+\omega',t')
$$
\begin{equation}\label{cr3}
 -\frac{e^{i(t'-t)(E_3-E_1-\omega')/\la^2}}{\la^2}
 \dl_{\ep_1,\ep_4}\dl(E_1-E_4)<g_{\ep_1},P_{E_1}g_{\ep_1}>{}
 N_{\ep_3,\ep_2,\la}(E_3,E_2,\omega+\omega',t).
\end{equation}
Notice that in the sense of distributions one has the limit
\begin{equation}\label{dllimit}
 \lim\limits_{\la\to 0}\frac{e^{i(t'-t)(E_1-E_2-\omega)/\la^2}}{\la^2}
  e^{it'(\omega-\omega')/\la^2}=2\pi\dl_{\omega,\omega'}
 \dl(t'-t)\dl(E_1-E_2-\omega)
\end{equation}
and, in the sense of distributions over the standard simplex
(see~\cite{ALV} for details) one has the limit
\begin{equation}\label{dlplimit}
 \lim\limits_{\la\to 0}\frac{e^{i(t'-t)(E_1-E_2-\omega)/\la^2}}{\la^2}
  e^{it'(\omega-\omega')/\la^2}=\dl_{\omega,\omega'}\dl_+(t'-t)\frac{1}{i(E_1-E_2-\omega-i0)}.
\end{equation}

The proof of the theorem follows by induction from the commutation
relations~(\ref{cr1})-(\ref{cr3}) using the limits~(\ref{dllimit})
and~(\ref{dlplimit}) and standard methods of the stochastic limit.
\qed


\section{The Master Space and the Associated White Noise}
\noindent In this section we construct a representation of the
limiting algebra~(\ref{crmf1})-(\ref{crmf3}).

Let $K$ be a vector space of finite rank operators acting on the
one particle Hilbert space ${\cal H}_1$ introduced in section~2,
with the property that, for any $\omega\in\Omega$ and for any $X,\
Y\in K$
$$
 <X,Y>_\omega:=\int dtTr\left(e^{-\beta H_1}X^*S_tYS^*_t\right)e^{-i\omega t}
$$
$$
 = 2\pi\int dETr\left(e^{-\beta H_1}X^*P_EYP_{E-\omega}\right)<\infty
$$
Because of our assumptions on ${\cal H}_1$, the space $K$ is non empty and $<\cdot,\cdot>_\om$
defines a prescalar product on $K$. We denote $\{K,<\cdot,\cdot>_\omega\}$ or
simply $K_\omega$ the Hilbert space with inner product $<\cdot,\cdot>_\omega$
obtained as completion of the quotient of $K$ by the zero $<\cdot,\cdot>_\omega$-norm
elements. Denoting $Y_t=S_tYS^*_t$ the free evolution of $Y$, one can rewrite the
inner product as
$$
 <X,Y>_\om=\int dtTr\left(e^{-\beta H_1}X^*Y_t\right)e^{-i\omega t}
$$

With these notations the representation space of the algebra~(\ref{crmf1})-(\ref{crmf3})
is the Fock space
$$
  \Gamma(L^2(\mathbb R_+)\otimes\bigoplus\limits_{\omega\in\Omega}K_\omega)\equiv
  \Gamma(\bigotimes\limits_{\om\in\Omega}L^2({\mathbb R}_+,K_\om))\equiv
  \bigotimes\limits_{\om\in\Omega}\Gamma(L^2({\mathbb R}_+,K_\om))
$$
where the infinite tensor product is referred to the vacuum vectors.
With these notations the master fields are realized as a family
of white noise operators $b_{t,\omega}(\cdot)$ which act
on $\Gamma(L^2(\mathbb R_+)\otimes\bigoplus\limits_{\omega\in\Omega}K_\omega)$
and satisfy the commutation relations
$$
 [b_{t,\omega}(X),b_{t',\omega'}^+(Y)]=\dl(t'-t)\dl_{\omega,\omega'}<X,Y>_\omega.
$$
Moreover each white noise operator $b_{t,\omega}(\cdot)$ acts as
an usual annihilation operator in $\Gamma(L^2(\mathbb R_+)\otimes
K_\omega)$ and as identity operator in other subspaces.

We will construct a representation of the algebra~(\ref{crmf1})-(\ref{crmf3})
in the Fock space defined above by the identification of
the operators $B_{\ep_1,\ep_2}(E_1,E_2,\omega,t)$
with the white noise operators
$$
 B_{\ep_1,\ep_2}(E_1,E_2,\omega,t)=b_{t,\omega}(|P_{E_1}g_{\ep_1}><P_{E_2}g_{\ep_2}|)
$$
The number operators will then have the form
$$
  N_{\ep_1,\ep_2}(E_1,E_2,\omega,t)=
$$
$$
  \sum\limits_{\ep\in\{0,1\}}\sum\limits_{\omega_1\in\Omega}
  \mu_{\ep}(E_1-\omega_1)b^+_{t,\omega_1}(|g_{\ep_1}><P_{E_1-\omega_1}g_{\ep}|)
  b_{t,\omega_1-\omega}(|P_{E_2}g_{\ep_2}><P_{E_1-\omega_1}g_{\ep}|)
$$
with
$$
 \mu_\ep(E):=\frac{1}{<g_\ep,P_EL^2g_\ep>}.
$$
One easily checks that these operators satisfy the algebra~(\ref{crmf1})-(\ref{crmf3}).

Let us introduce, for simplicity of calculations, the operators
$$
 B_{\ep_1,\ep_2}(E,\omega,t):=\int dE'B_{\ep_1,\ep_2}(E',E,\omega,t)
$$
$$
 N_{\ep_1,\ep_2}(\omega,t):=\int dE_1dE_2N_{\ep_1,\ep_2}(E_1,E_2,\omega,t)=
  \int dEN_{\ep_1,\ep_2}(E,\om,t)
$$
with
$$
  N_{\ep_1,\ep_2}(E,\omega,t)=\sum\limits_{\ep\in\{0,1\}}\sum\limits_{\omega_1\in\Omega}
  \mu_{\ep}(E)B^+_{\ep_1,\ep}(E,\omega_1,t)
  B_{\ep_2,\ep}(E,\omega_1-\omega,t)
$$

The operators $B_{\ep_1,\ep_2}(E,\omega,t)$,
$B^+_{\ep_1,\ep_2}(E,\omega,t)$ and $N_{\ep_1,\ep_2}(\omega,t)$
satisfy the (causal) commutation relations:
$$
 \vphantom{\e}
 [B_{\ep_1,\ep_2}(E,\omega,t),B^+_{\ep_3,\ep_4}(E',\omega',t')]
$$
\begin{equation}\label{comrel1}\vphantom{\e}
 =\dl_+(t'-t)\dl_{\ep_1,\ep_3}\dl_{\ep_2,\ep_4}\dl_{\omega,\omega'}\dl(E-E')
 \ga_{\ep_1}(E+\omega)\mu^{-1}_{\ep_2}(E)
\end{equation}
\begin{equation}\label{comrel2}
 \vphantom{\e}
 [B_{\ep_1,\ep_2}(E,\omega,t),N_{\ep_3,\ep_4}(\omega',t')]=
 \dl_+(t'-t)\dl_{\ep_1,\ep_3}\gamma_{\ep_1}(E+\omega)B_{\ep_4,\ep_2}(E,\omega-\omega',t').
\end{equation}

In these notations the limiting white noise Hamiltonian
acts on $\Gamma(L^2(\mathbb R_+)\otimes\bigoplus\limits_{\omega\in\Omega}K_\omega)$ as
$$
 H(t)=\sum\limits_{\omega\in B}D_\omega\otimes\Bigl\{
 N_{0,1}(\omega,t)+\int dE\Bigl[B_{1,0}(E,-\omega,t)+
 B_{0,1}^+(E,\omega,t)\Bigr]\Bigr\}
$$
$$
 +\sum\limits_{\omega\in B}D^+_{-\omega}\otimes\Bigl\{
 N_{1,0}(\omega,t)+\int dE\Bigl[B_{0,1}(E,-\omega,t)+
 B_{1,0}^+(E,\omega,t)\Bigr]\Bigr\}.
$$


\section{White Noise Stochastic Schr\"odinger Equation}
\noindent The results of the preceding section allow us to write
the white noise Schr\"odinger equation for the evolution operator
in the stochastic limit
$$
 \partial_tU_t=-iH(t)U_t=-i\sum\limits_{\omega\in B}\Bigl\{D_\omega\otimes\Bigl[
 N_{0,1}(\omega,t)+\int dE\Bigl(B_{1,0}(E,-\omega,t)+
 B_{0,1}^+(E,\omega,t)\Bigr)\Bigr]
$$
\begin{equation}\label{equ}
 +D^+_{-\omega}\otimes\Bigl[N_{1,0}(\omega,t)+\int dE\Bigl(B_{0,1}(E,-\omega,t)+
 B_{1,0}^+(E,\omega,t)\Bigr)\Bigr]\Bigr\}U_t
\end{equation}

Following the general theory of white noise equations, in order
to give a precise meaning to this equation we will rewrite it
in the causally normally ordered form
in which all annihilators are on the right hand side of the
evolution operator and all creators  on the left hand side.
After this procedure we obtain a quantum stochastic differential
equation (QSDE) in the sense of Hudson and Parthasarathy~\cite{HP}.

For each $\omega\in\Omega\backslash{\rm B}$ define $D_\omega=0$.
Then define for any $\om,\om'\in\Omega$ the operators $T^0_{\omega,\omega'}(E)$
and $T^1_{\omega,\omega'}(E)$ by
$$
 T^0_{\omega,\omega'}(E):=\sum\limits_{\omega''}\gamma_0(E+\omega)
 \gamma_1(E-\omega'')D_{\omega''+\omega}D^+_{\omega''+\omega'}
$$
$$
 T^1_{\omega,\omega'}(E):=\sum\limits_{\omega''}\gamma_1(E+\omega)
 \gamma_0(E+\omega'')D^+_{\omega''-\omega}D_{\omega''-\omega'}
$$
We will also denote for each $\ep=0,1$
$$
 (1+T_\ep)_{\omega,\omega'}(E)=\dl_{\omega,\omega'}+T^\ep_{\omega,\omega'}(E)
$$

The following lemma plays an important role in the derivation of
the normally ordered white noise equation.

\begin{lemma}\label{normordu}
In the above notations one has
$$
 \sum\limits_{\omega'}(1+T_0)_{\omega,\omega'}(E)B_{0,0}(E,\omega',t)U_t=U_tB_{0,0}(E,\omega,t)
 -i\sum\limits_{\omega'}D_{\om-\omega'}\gamma_0(E+\omega)
$$
\begin{equation}\label{B00}
 \times U_tB_{1,0}(E,\omega',t)-\sum\limits_{\omega'}D_{\omega'+\omega}D^+_{\omega'}\mu^{-1}_0(E)
 \gamma_0(E+\omega)\gamma_1(E-\omega')U_t
\end{equation}
$$
 \sum\limits_{\omega'}(1+T_0)_{\omega,\omega'}(E)B_{0,1}(E,\omega',t)U_t=
 U_tB_{0,1}(E,\omega,t)
 -i\sum\limits_{\omega'}D_{\om-\omega'}\gamma_0(E+\omega)
$$
\begin{equation}\label{B01}\vphantom{\sum\limits_{\om'}}
 \times U_tB_{1,1}(E,\omega',t)-iD_{\omega}\mu^{-1}_1(E)\gamma_0(E+\omega)U_t
\end{equation}
$$
 \sum\limits_{\omega'}(1+T_1)_{\omega,\omega'}(E)B_{1,1}(E,\omega',t)U_t=U_tB_{1,1}(E,\omega,t)
 -i\sum\limits_{\omega'}D^+_{\omega'-\om}\gamma_1(E+\omega)
$$
\begin{equation}\label{B11}
 \times U_tB_{0,1}(E,\omega',t)-\sum\limits_{\omega'}D^+_{\omega'-\omega}D_{\omega'}\mu^{-1}_1(E)
 \gamma_1(E+\omega)\gamma_0(E+\omega')U_t
\end{equation}
$$
 \sum\limits_{\omega'}(1+T_1)_{\omega,\omega'}(E)B_{1,0}(E,\omega',t)U_t=U_tB_{1,0}(E,\omega,t)
 -i\sum\limits_{\omega'}D^+_{\omega'-\om}\gamma_1(E+\omega)
$$
\begin{equation}\label{B10}\vphantom{\sum\limits_{\om'}}
 \times U_tB_{0,0}(E,\omega',t)-iD^+_{-\omega}\mu^{-1}_0(E)\gamma_1(E+\omega)U_t
\end{equation}
\end{lemma}
{\bf Proof.} It is clear that
$$
 B_{\ep_1,\ep_2}(E,\omega,t)U_t=[B_{\ep_1,\ep_2}(E,\omega,t),U_t]+
 U_tB_{\ep_1,\ep_2}(E,\omega,t)
$$

Using the integral form of equation~(\ref{equ}) for the evolution
operator one gets
$$
 [B_{\ep_1,\ep_2}(E,\omega,t),U_t]=-i\int\limits_0^t dt_1\Bigl([B_{\ep_1,\ep_2}(E,\omega,t),H(t_1)]U_{t_1}
$$
\begin{equation}\label{commBU}
+H({t_1})[B_{\ep_1,\ep_2}(E,\omega,t),U_{t_1}]\Bigr)
\end{equation}
Notice that due to the time consecutive principle (see~\cite{ALV} for details)
one has for $t>t_1$
$$
 [B_{\ep_1,\ep_2}(E,\omega,t),U_{t_1}]=0.
$$

Using the causal commutation relations~(\ref{comrel1}),~(\ref{comrel2})
one can compute the causal commutator $[B_{\ep_1,\ep_2}(E,\omega,t),H(t_1)]$.
After substitution of this commutator in~(\ref{commBU}) one gets
$$
 [B_{\ep_1,\ep_2}(E,\omega,t),U_t]=
$$
$$
 -i\dl_{0,\ep_1}\gamma_0(E+\omega)\Bigl(\sum\limits_{\omega'}
 D_{\omega'}B_{1,\ep_2}(E,\omega-\omega',t)U_t+
 \dl_{1,\ep_2}D_{\omega}\mu^{-1}_1(E)U_t\Bigr)
$$
$$
 -i\dl_{1,\ep_1}\gamma_1(E+\omega)\Bigl(\sum\limits_{\omega'}
 D^+_{-\omega'}B_{0,\ep_2}(E,\omega-\omega',t)U_t+
 \dl_{0,\ep_2}D^+_{-\omega}\mu^{-1}_0(E)U_t\Bigr)
$$
From this it follows that
\begin{equation}\label{B00U}
 B_{0,0}(E,\omega,t)U_t=U_tB_{0,0}(E,\omega,t)
 -i\gamma_0(E+\omega)\sum\limits_{\omega'}
 D_{\omega'}B_{1,0}(E,\omega-\omega',t)U_t
\end{equation}
$$
 \vphantom{\sum\limits_{\om_n}}
 B_{1,0}(E,\omega,t)U_t=U_tB_{1,0}(E,\omega,t)
 -i\gamma_1(E+\omega)
$$
\begin{equation}\label{B10U}
 \times\Bigl(\sum\limits_{\omega'}
 D^+_{-\omega'}B_{0,0}(E,\omega-\omega',t)U_t+
 D^+_{-\omega}\mu^{-1}_0(E)U_t\Bigr)
\end{equation}

After substitution of~(\ref{B10U}) in~(\ref{B00U}) one gets
$$
 B_{0,0}(E,\omega,t)U_t=-\sum\limits_{\omega'}\sum\limits_{\omega''}
 D_{\omega'}D^+_{-\omega''}
 \gamma_0(E+\omega)\gamma_1(E+\omega-\omega')B_{0,0}(E,\omega-\omega'-\omega'',t)U_t
$$
$$
 +U_tB_{0,0}(E,\omega,t)-i\gamma_0(E+\omega)\sum\limits_{\omega'}
 D_{\omega'}U_tB_{1,0}(E,\omega-\omega',t)
$$
$$
 -\sum\limits_{\omega'}D_{\omega'}D^+_{\omega'-\omega}
 \gamma_0(E+\omega)\gamma_1(E+\omega-\omega')\mu^{-1}_0(E)U_t
$$
Now changing the summation index in the double sum,
$\tilde\omega''=\omega-\omega'-\omega''$, using the definition of
$T^0_{\om,\om'}(E)$, and bringing the double sum to the left hand
side of the equality, one obtains~(\ref{B00}). The derivation
of~(\ref{B01})-(\ref{B10}) can be done in a similar way.\, \qed


\section{The Normally Ordered Equation}
\noindent In this section we will bring equation~(\ref{equ}) to
the normally ordered form. For this goal we will express terms
like $B_{\ep_1,\ep_2}(E,\omega,t)U_t$ in the RHS of
equation~(\ref{equ}) in a form in which the annihilation operators
are on the RHS of the evolution operator. This form is based on
equations~(\ref{B00})-(\ref{B10}) of Lemma~1, which we will solve
with respect to $B_{\ep_1,\ep_2}(E,\omega,t)U_t$.

Suppose that for each $\ep=0,1$ the operators
$(1+T_\ep)_{\omega,\omega'}(E)$ are invertible so that there exist
the operators $(1+T_\ep)^{-1}_{\omega,\omega''}(E)$ with the
propertie
$$
 \sum\limits_{\omega''}(1+T_\ep)^{-1}_{\omega,\omega''}(E)
(1+T_\ep)_{\omega'',\omega'}(E)=\dl_{\omega,\omega'}
$$
and that these operators are given by the convergent series
$$
 (1+T_\ep)^{-1}_{\omega,\omega'}(E)=\dl_{\omega,\omega'}+\sum\limits_{n=1}^\infty(-1)^n
$$
\begin{equation}\label{Tseries}
 \times\sum\limits_{\om_1,\dots,\om_{n-1}}T^\ep_{\om,\om_1}(E)T^\ep_{\omega_1,\omega_2}(E)
 \dots T^\ep_{\omega_{n-1},\omega'}(E)
\end{equation}
Detailed investigation of conditions under which this series
converges will be done in a future paper.

Then let us define
$$
 R^{0,1}_{\om,\om'}(E)=-i\sum\limits_{\om_1}D_{\om-\om_1}(1+T_1)^{-1}_{\om_1,\om'}(E)
$$
$$
 R^{1,0}_{\om,\om'}(E)=-i\sum\limits_{\om_1}D^+_{\om_1-\om}(1+T_0)^{-1}_{\om_1,\om'}(E)
$$
$$
 R^{0,0}_{\om,\om'}(E)=-\sum\limits_{\om_1,\om_2}D_{\om-\om_1}(1+T_1)^{-1}_{\om_1,\om_2}(E)
  D^+_{\om'-\om_2}\gamma_1(E+\om_2)
$$
$$
 R^{1,1}_{\om,\om'}(E)=-\sum\limits_{\om_1,\om_2}D^+_{\om_1-\om}(1+T_0)^{-1}_{\om_1,\om_2}(E)
  D_{\om_2-\om'}\gamma_0(E+\om_2)
$$
With these notations the normally ordered form of the
equation~(\ref{equ}) is given by the following theorem

\begin{theorem}
The normally ordered form of equation~(\ref{equ}) is
$$
 \partial_tU_t=\sum\limits_{\ep_1,\ep_2}\int dE\Bigl[\, \sum\limits_{\om,\om'}
  R^{\ep_1,\ep_2}_{\om,\om'}(E)\sum\limits_\ep\mu_\ep(E)B^+_{\ep_1,\ep}(E,\om,t)
   U_tB_{\ep_2,\ep}(E,\omega',t)
$$
$$
 +\sum\limits_\om\Bigl(R^{\ep_1,\ep_2}_{\om,0}(E)B^+_{\ep_1,\ep_2}(E,\om,t)U_t+
  R^{\ep_2,\ep_1}_{0,\om}(E)U_tB_{\ep_1,\ep_2}(E,\om,t)\Bigr)
$$
\begin{equation}\label{normordeq}
 +R^{\ep_1,\ep_2}_{0,0}(E)<g_{\ep_1},P_Ee^{-\beta H_1}g_{\ep_2}>U_t\Bigr]
\end{equation}
\end{theorem}

\noindent {\bf Proof.} Using the inverse operators
$(1+T_\ep)^{-1}_{\om,\om'}(E)$ in
equations~(\ref{B00})-(\ref{B10}) one can express the products
$B_{\ep_1,\ep_2}(E,\om,t)U_t$ in terms of the products
$U_tB_{\ep'_1,\ep'_2}(E,\om',t)$. For example:
$$
 B_{0,0}(E,\omega,t)U_t=
 \sum\limits_{\omega'}(1+T_0)^{-1}_{\omega,\omega'}(E)\Bigl[U_tB_{0,0}(E,\omega',t)
 -i\sum\limits_{\omega''}D_{\om'-\omega''}\gamma_0(E+\omega')
$$
$$
 \times U_tB_{1,0}(E,\omega'',t)-\sum\limits_{\omega''}D_{\omega''+\omega'}D^+_{\omega''}<g_0,P_EL^2g_0>
 \gamma_0(E+\omega')\gamma_1(E-\omega'')U_t\Bigr]
$$
and similarly for the other terms. Then after substitution of
these expressions in~(\ref{equ}) one obtains~(\ref{normordeq}).\,
\qed

It is known~\cite{ALV} that any normally ordered white noise
equation is equivalent to a quantum stochastic differential equation.
In particular equation~(\ref{normordeq}) is equivalent to
the quantum stochastic differential equation for the evolution operator:
$$
 dU_t=\sum\limits_{\ep_1,\ep_2}\int dE\Biggl[\, \sum\limits_{\om,\om'}
 R^{\ep_1,\ep_2}_{\om,\om'}(E)dN_t(Z^{\ep_1,\ep_2}_{\om,\om'}(E))
$$
$$
 +\sum\limits_\om\Bigl(R^{\ep_1,\ep_2}_{\om,0}(E)dB^+_t((|g_{\ep_1}><P_Eg_{\ep_2}|)_\om)+
  R^{\ep_2,\ep_1}_{0,\om}(E)dB_t((|g_{\ep_1}><P_Eg_{\ep_2}|)_\om)\Bigr)
$$
\begin{equation}\label{QSDE}
 +R^{\ep_1,\ep_2}_{0,0}(E)<g_{\ep_1},P_Ee^{-\beta H_1}g_{\ep_2}>dt\Biggr]U_t
\end{equation}
Here we denote by $(|f><g|)_\om$ the element of
$\bigoplus\limits_{\om\in\Omega}K_\om$ which belongs to $K_\om$
subspace. Moreover,
$$
 Z^{\ep_1,\ep_2}_{\om,\om'}(E): \bigoplus\limits_{\om\in\Omega}K_\om\to
\bigoplus\limits_{\om\in\Omega}K_\om
$$
and acts on an element $X\in K_{\om''}$ as
$$
 Z^{\ep_1,\ep_2}_{\om,\om'}(E)X=\dl_{\om',\om''}\sum\limits_\ep\mu_\ep(E)
 <(|g_{\ep_2}><P_Eg_\ep|)_{\om'}, X>_{\om'}(|g_{\ep_1}><P_Eg_\ep|)_\om\in K_\om
$$


\section{Connection with Scattering Theory}
\noindent Let us show that the evolution operator in the LDL is
directly related with the (1-particle) $T$-operator describing the
scattering of the system on one particle of the reservoir. Notice
that the mean value of the evolution operator~(\ref{EvOp}) with
respect to the state~(\ref{state}) in the low density limit is
equal to the vacuum mean value of the solution of the
QSDE~(\ref{QSDE})
\begin{equation}\label{expvalue}
 \lim\limits_{\xi\to 0}\varphi^\xi\Bigl(U(t/\xi)\Bigr)=<U_t>_{vac}=e^{-\Gamma t}
\end{equation}
where, since we average only over the reservoir degrees of freedom,
the drift term $ \Gamma $ is an operator acting in the system
Hilbert space $ {\cal H}_S $ as
$$
 \Gamma =-\sum\limits_{\ep=0,1}\int dE
 R^{\ep,\ep}_{0,0}(E)<g_\ep,P_Ee^{-\beta H_1}g_\ep>
$$
with $R^{\ep,\ep}_{0,0}(E)$ given at the beginning of section~7.

Let us remind the definition of $T$-operator. For the interaction
of scattering type the closed subspace of $ {\cal H}_S\otimes
\Gamma({\cal H}_1) $ generated by vectors of the form $ u\otimes
A^+(f)\Phi $\enskip ($ u\in {\cal H}_S $, $ f\in {\cal H}_1 $ and
$\Phi\in\Gamma({\cal H}_1)$ is the vacuum vector) which is
naturally isomorphic to $ {\cal H}_S\otimes {\cal H}_1 $, is
globally invariant under the time evolution operator $
\exp[i(H_S\otimes 1+1\otimes H_R+V)t] $. Explicitly the
restriction of the time evolution operator to this subspace is
given by
$$
 \exp[i(H_S\otimes 1+1\otimes H_1+V_1)t]\in{\cal B}({\cal H}_S\otimes {\cal H}_1)
$$
with
\begin{equation}\label{defV}
 V_1=\sum\limits_{\ep=0,1}D_\ep\otimes |g_\ep><g_{1-\ep}|
\end{equation}

The 1-particle M\o ller wave operators are defined by:
\begin{equation}\label{Moller}
 \Omega_{\pm}=s-\lim\limits_{t\to\pm\infty}\exp[-i(H_S\otimes 1+1\otimes H_1+V_1)t]
 \exp[i(H_S\otimes 1+1\otimes H_1)t]
\end{equation}
and the 1-particle $ T $-operator is defined by:
\begin{equation}\label{defT}
 T=V_1\Omega_+
\end{equation}
From~(\ref{defV}) and~(\ref{Moller}) it follows that
$$
 \Omega_{\pm}=s-\lim\limits_{t\to\pm\infty}U^{(1)}_t
$$
where $ U_t^{(1)} $ is the solution of
$$
 \partial_tU_t^{(1)}=-iU_t^{(1)}V_1(t),\quad U_0^{(1)}=1.
$$
Here
\begin{equation}\label{V1(t)}
 V_1(t)=\sum\limits_\om D_\om e^{it\om}\otimes |S_{-t}g_0><S_{-t}g_1|+
 \sum\limits_\om D^+_\om e^{-it\om}\otimes |S_{-t}g_1><S_{-t}g_0|
\end{equation}
We will show in Appendix that
\begin{equation}\label{Top2}
 T=i\sum\limits_{\ep,\ep'=0,1}\int dE\sum\limits_\om R^{\ep,\ep'}_{\om,0}(E)
  \otimes |g_\ep><g_{\ep'}|P_E.
\end{equation}
From this formula it follows (see Appendix) that the drift
term $\Gamma$ is connected with the $T$-operator by the equality
\begin{equation}\label{drift}
 \Gamma=i\sum\limits_kP_k\left(Tr_{{\cal H}_1}e^{-\beta H_1}T\right)P_k
\end{equation}
This formula means that the drift term is the diagonal part (in
the sense of the spectral projections of the system Hamiltonian)
of the partial expectation of the $T$-operator in the 1-particle
reservoir Gibbs state. In particular, under the RWA
assumption~(\ref{rwa}) with $\om_0 = 0$ the expression for the
drift term has the form
$$
 \Gamma=iTr_{{\cal H}_1}e^{-\beta H_1}T
$$
Formula~(\ref{drift}) has important implications for the master
equation which will be discussed in a future paper.


\section{The Langevin Equation}
\noindent Using the stochastic golden rule we can find the
Langevin equation, which is the limit of the Heisenberg evolution
in interaction representation, of any observable $X=X_S\otimes
1_R$ of the system. The Langevin equation is the equation
satisfied by the stochastic flow $j_t$, defined by
$$
 j_t(X)\equiv X_t :=U^*_tXU_t
$$
where $U_t$ satisfies equation~(\ref{QSDE}). To derive the
Langevin equation we may apply the stochastic golden rule,
described in previous sections. As a result we get in terms of the
white noise operators
$$
 \dot X_t=\sum\limits_{\ep_1,\ep_2}\int dE\Biggl[\, \sum\limits_{\om_1,\om_2}
 \sum\limits_\ep\mu_\ep(E)B^+_{\ep_1,\ep}(E,\om_1,t)
 U^*_t\Theta^{\ep_1,\ep_2}_{\om_1,\om_2}(X)U_tB_{\ep_2,\ep}(E,\om_2,t)
$$
$$
 +\sum\limits_\om\Bigl(B^+_{\ep_1,\ep_2}(E,\om,t)U^*_t\Theta^{\ep_1,\ep_2}_{\om,0}(X)U_t+
 U^*_t\Theta^{\ep_2,\ep_1}_{0,\om}(X)U_tB_{\ep_1,\ep_2}(E,\om,t)\Bigr)
$$
\begin{equation}\label{Langevin}
 +<g_{\ep_1},P_Ee^{-\beta H_1}g_{\ep_2}>U^*_t\Theta^{\ep_1,\ep_2}_{0,0}(X)U_t\Biggr]
\end{equation}
with the maps
$$
 \Theta^{\ep_1,\ep_2}_{\om_1,\om_2}(X):=XR^{\ep_1,\ep_2}_{\om_1,\om_2}(E)+
 R^{+ \ep_2,\ep_1}_{\om_2,\om_1}(E)X+2\sum\limits_{\ep,\om}{\rm Re}\gamma_\ep(E+\om)
 R^{+\ep,\ep_1}_{\om,\om_1}(E)XR^{\ep,\ep_2}_{\om,\om_2}(E)
$$
The equation~(\ref{Langevin}) can be rewritten in terms of the
stochastic differentials as
$$
 dj_t(X)=j_t\circ\sum\limits_{\ep_1,\ep_2}\int dE\Biggl[\, \sum\limits_{\om_1,\om_2}
 \Theta^{\ep_1,\ep_2}_{\om_1,\om_2}(X)dN_t(Z^{\ep_1,\ep_2}_{\om_1,\om_2}(E))
$$
$$
 +\sum\limits_\om\Bigl(\Theta^{\ep_1,\ep_2}_{\om,0}(E)dB^+_t((|g_{\ep_1}><P_Eg_{\ep_2}|)_\om)+
 \Theta^{\ep_2,\ep_1}_{0,\om}(E)dB_t((|g_{\ep_1}><P_Eg_{\ep_2}|)_\om)\Bigr)\Biggr]
$$
\begin{equation}\label{langevin}
 +j_t\circ\Theta_0(X)dt
\end{equation}
Here
$$
 \Theta_0(X):=\sum\limits_{\ep}\int dE\mu^{-1}_{\ep}(E)\Bigl[
 XR^{\ep,\ep}_{0,0}(E)+
 R^{+ \ep,\ep}_{0,0}(E)X
$$
\begin{equation}\label{QMG}
 +2\sum\limits_{\ep',\om}{\rm Re}\gamma_{\ep'}(E+\om)
 R^{+ \ep',\ep}_{\om,0}(E)XR^{\ep',\ep}_{\om,0}(E)\Bigr]
\end{equation}
is a quantum Markovian generator. The structure map $\Theta_0(X)$
has the standard form of the generator of a master
equation~\cite{lindblad}
$$
 \Theta_0(X)=\Psi(X)-\frac{1}{2}\{\Psi(1),X\}+i[H,X]
$$
where
$$
 \Psi(X):=2\sum\limits_{\ep}\int dE\mu^{-1}_{\ep}(E)\sum\limits_{\ep',\om}
 {\rm Re}\gamma_{\ep'}(E+\om)
 R^{+ \ep',\ep}_{\om,0}(E)XR^{\ep',\ep}_{\om,0}(E)
$$
is a completely positive map and
$$
 H:=\sum\limits_{\ep}\int dE\mu^{-1}_{\ep}(E)\frac{
 R^{+ \ep,\ep}_{0,0}(E)-R^{\ep,\ep}_{0,0}(E)}{2i}
$$
is selfadjoint.


\section{Conclusions}\noindent
An important problem in theory of open quantum systems is a rigorous derivation of a quantum
Boltzmann equation. In the present paper the quantum model of the test particle interacting with
the Bose gas has been considered. For this model we have presented a rigorous derivation of the
quantum Langevin and the quantum master equation.

We developed the stochastic limit method for the low density case. The procedure of the deduction
of a (unitary) evolution of the compound system in the limit of long time and small density of
particles of the gas was developed. This procedure is being called the stochastic golden rule for
the low density limit. The limiting evolution is directly expressed in terms of the 1-particle
$T$--operator describing scattering of the test particle on one particle of the reservoir. After
that we obtain the quantum Langevin equation. This equation includes not only the system but also
the reservoir dynamics. The master equation can be obtained after averaging of this equation over
the equilibrium state of the reservoir. We find the generator of this master equation which
describes the reduced evolution of the test particle. We show that this generator has the Lindblad
form of most general generators of completely positive semigroups.

We considered the situation when the temperature of the reservoir is high enough so that no
condensate is present. It is an interesting important problem to generalize the stochastic limit
method to the case that Bose condensate is present.


\renewcommand{\theequation}{\Alph{appendixc}.\arabic{equation}}
\appendix

\noindent Let us first derive an explicit formula for the
$T$-operator~(\ref{Top2}).

The perturbation expansion for $\Omega_+$ is
$$
 \Omega_+=\sum\limits_{n=0}^\infty (-i)^n\int_0^\infty dt_1\dots\int_0^{t_{n-1}}
 dt_n V_1(t_n)\dots V_1(t_1)
$$
This expansion induces the following expansion for the $T$-operator:
$$
 T=\sum\limits_{n=1}^\infty T_n
$$
with
$$
 T_{n+1}=(-i)^n \int_0^\infty dt_1\dots\int_0^{t_{n-1}}
 dt_n V_1V_1(t_n)\dots V_1(t_1)
$$
and $V_1(t)$ is given by~(\ref{V1(t)}).

By direct calculations one can prove that
$$
 T_{2n}=i\int dE\left(T^{00}_{2n}(E)|g_0><P_Eg_0|+T^{11}_{2n}(E)|g_1><P_Eg_1|\right)
$$
$$
 T_{2n+1}=i\int dE\left(T^{01}_{2n+1}(E)|g_0><P_Eg_1|+T^{10}_{2n+1}(E)|g_1><P_Eg_0|\right)
$$
with
$$
 T^{00}_{2n}(E)=(-1)^n\sum\limits_{\om,\om_1\dots\omega_{2n-1}}
 D_\om D^+_{\om_1}\dots D_{\om_{2n-2}}D^+_{\om_{2n-1}}
 \gamma_1(E-\om_{2n-1})
$$
$$
 \vphantom{\sum\limits_{\om_n}}
 \times\gamma_0(E-\om_{2n-1}+\om_{2n-2})\dots
 \gamma_0(E-\om_{2n-1}+\dots+\omega_2)
 \gamma_1(E-\om_{2n-1}+\dots+\omega_2-\omega_1)
$$
$$
 T^{11}_{2n}(E)=(-1)^n\sum\limits_{\om,\om_1\dots\omega_{2n-1}}
 D^+_\omega D_{\omega_1}\dots D^+_{\omega_{2n-2}}D_{\omega_{2n-1}}
 \gamma_0(E+\omega_{2n-1})
$$
$$
 \vphantom{\sum\limits_{\om_n}}
 \times\gamma_1(E+\omega_{2n-1}-\omega_{2n-2})\dots
 \gamma_1(E+\omega_{2n-1}-\dots-\omega_2)
 \gamma_0(E+\omega_{2n-1}-\dots-\omega_2+\omega_1)
$$
$$
 T^{01}_{2n+1}(E)=-i(-1)^n\sum\limits_{\om,\om_1\dots\om_{2n}}
 D_\omega D^+_{\omega_1}\dots D^+_{\omega_{2n-1}}D_{\omega_{2n}}
 \gamma_0(E+\omega_{2n})
$$
$$
 \vphantom{\sum\limits_{\om_n}}
 \times\gamma_1(E+\omega_{2n}-\omega_{2n-1})\dots
 \gamma_0(E+\omega_{2n}-\dots+\omega_2)
 \gamma_1(E+\omega_{2n}-\dots+\omega_2-\omega_1)
$$
$$
 T^{10}_{2n+1}(E)=-i(-1)^n\sum\limits_{\om,\om_1\dots\om_{2n}}
 D^+_\omega D_{\omega_1}\dots D_{\omega_{2n-1}}D^+_{\omega_{2n}}
 \gamma_1(E-\omega_{2n})
$$
$$
 \vphantom{\sum\limits_{\om_n}}
 \times\gamma_0(E-\omega_{2n}+\omega_{2n-1})\dots
 \gamma_1(E-\omega_{2n}+\dots-\omega_2)
 \gamma_0(E-\omega_{2n}+\dots-\omega_2+\omega_1)
$$

Let us show, for example, that
\begin{equation}
 \sum\limits_\om R^{1,0}_{\om,0}(E)=\sum\limits_{n=0}^\infty T^{10}_{2n+1}(E) \label{R10}
\end{equation}
In fact,
$$
 \sum\limits_\om R^{1,0}_{\om,0}(E)=-i\sum\limits_{\om,\om_1}D^+_\om(1+T_0)^{-1}_{\om_1,0}(E)=
$$
$$
 -i\sum\limits_{\om,\om_1}D^+_\om\dl_{\om_1,0}-i\sum\limits_{n=1}^\infty (-1)^n
  \sum\limits_{\om,\om_1\dots\om_n}D^+_\om T^0_{\om_1,\om_2}(E)T^0_{\om_2,\om_3}(E)\dots
  T^0_{\om_n,0}(E)
$$
One has
$$
 (-1)^n\sum\limits_{\om,\om_1\dots\om_n}D^+_\om T^0_{\om_1,\om_2}(E)T^0_{\om_2,\om_3}(E)\dots
  T^0_{\om_n,0}(E)=
$$
$$
 (-1)^n\sum\limits_{\om,\om_1,\om_3\dots\om_{2n-1}}D^+_\om   T^0_{\om_1,\om_3}(E)T^0_{\om_3,\om_5}(E)\dots T^0_{\om_{2p-1},\om_{2p+1}}(E)\dots
 T^0_{\om_{2n-1},0}(E)=
$$
$$
 (-1)^n\sum\limits_{\om,\om_1,\om_2\dots\om_{2n}}D^+_\om D_{\om_2+\om_1}D^+_{\om_2+\om_3}\dots
 D_{\om_{2p}+\om_{2p-1}}D^+_{\om_{2p}+\om_{2p+1}}\dots D_{\om_{2n}+\om_{2n-1}}D^+_{\om_{2n}}
$$
$$
 \vphantom{\sum\limits_{\om_n}}
 \times\gamma_0(E+\om_1)\gamma_1(E-\om_2)\dots\gamma_0(E+\om_{2p-1})\gamma_1(E-\om_{2p})\dots
 \gamma_0(E+\om_{2n-1})\gamma_1(E-\om_{2n})
$$
Now let us make a change of summation index $\om_{2n-1}\to\om_{2n-1}-\om_{2n}$, then
$\om_{2n-2}\to\om_{2n-2}-\om_{2n-1}+\om_{2n}$,~... and finally $\om_1\to\om_1-\om_2+\dots+\om_{2n-1}-\om_{2n}$. After this change of summation indices we get for the RHS of the previous equality
$$
 (-1)^n\sum\limits_{\om,\om_1\dots\om_{2n}}D^+_\om D_{\om_1}D^+_{\om_2}\dots
 D_{\om_{2p-1}}D^+_{\om_{2p}}\dots D_{\om_{2n-1}}D^+_{\om_{2n}}
 \gamma_0(E+\om_1-\om_2+\dots-\om_{2n})
$$
$$
 \times\gamma_1(E-\om_2+\dots-\om_{2n})\dots
 \gamma_0(E+\om_{2n-1}-\om_{2n})\gamma_1(E-\om_{2n})\equiv T^{01}_{2n+1}(E)
$$
and from this~(\ref{R10}) follows. One can prove in the similar way that
$$
 \sum\limits_\om R^{0,1}_{\om,0}(E)=\sum\limits_{n=0}^\infty T^{01}_{2n+1}(E),\quad
 \sum\limits_\om R^{\ep,\ep}_{\om,0}(E)=\sum\limits_{n=1}^\infty T^{\ep\,\ep}_{2n}(E),\quad
 \ep=0,1
$$
Therefore the $T$-operator is given by formula~(\ref{Top2}).

Now let us prove the relation~(\ref{drift}) beetwen the drift term and
the $T$-operator. Since $g_0$ and $g_1$ are mutually ortogonal one has
$$
 iTr_{{\cal H}_1}e^{-\beta H_1}T=-\sum\limits_{\ep=0,1}\int dE
 \sum\limits_\om    R^{\ep,\ep}_{\om,0}(E)<g_\ep,P_Ee^{-\beta H_1}g_\ep>
$$
Then by expanding the operators $(1+T_\ep)^{-1}_{\om,\om'}$ in
the series~(\ref{Tseries}) one can show that for $\om\ne 0$
$$
 \sum\limits_kP_kR^{\ep,\ep}_{\om,0}P_k=0
$$
and
$$
 \sum\limits_kP_kR^{\ep,\ep}_{0,0}P_k=R^{\ep,\ep}_{0,0}
$$
From this~(\ref{drift}) follows.


\nonumsection{Acknowledgment}\noindent

This work is partially supported by the INTAS 99-0590 for L.~A. and I.~V. and by the INTAS
01/1-200 for A.~P. and also by the RFFI 02-01-01084 and the grant of the leading scientific school
00-15-96073.

\nonumsection{References} 

\end{document}